# Shorted Micro-Waveguide Array for High Optical Transparency and Superior Electromagnetic Shielding in Ultra-Wideband Frequency Spectrum

Yuanlong Liang[1], Xianjun Huang[1]*, Jisheng Pan[1], Wencong Liu[1], Kui Wen[1], Duocai Zhai[1], Peng Shang[2]*, and Peiguo Liu[1]

[1]College of Electronic Science and Technology, National University of Defense Technology Changsha, 410073, China

[2]GRINM Guojing Advanced Materials Co., Ltd, Beijing, 100088, China

*To whom correspondence and requests for materials should be addressed:

huangxianjun@nudt.edu.cn; shangpeng163@163.com

## Abstract

While functional materials with both light transmitting and electromagnetic shielding are highly desirable, only very few of them meet the stringent electromagnetic interference (EMI) shielding criteria for optoelectronic systems. Here, a design strategy of shorted micro-waveguides (SMWs) array to decouple the light transmission and EMI shielding is proposed and experimentally demonstrated. The array of SMWs, consisting of cut-off metallic micro-waveguides and shorting indium tin oxide (ITO) continuous conductive film, exhibits high optical transmittance of 90.4% and superior EMI shielding effectiveness of 62.2 dB on average over ultra-wide frequency spectrum (0.2–1.3 GHz & 1.7–18 GHz). Compared to previously reported works, an improvement of 19 dB in average shielding effectiveness has been achieved under the same level of light transmission, and the shielding frequency spectrum has been significantly expanded. The working principle has been explained in depth and factors influencing the performance have been investigated for design optimization.

## Introduction

The electromagnetic (EM) environment becomes increasingly complex as wireless technology advances. Due to the intensification of electromagnetic radiation and the expansion of electromagnetic spectrum, severe electromagnetic interference (EMI) risks have arisen to electronic devices as well as human health[1-5]. Especially for most sensitive components in optoelectronic systems, the wideband high intensity radiated fields (HIRF) can be coupled into electronic equipment through optical windows due to their lack of electromagnetic reinforcement, resulting in detrimental influences on the stability and security of entire systems[6,7]. It is becoming increasingly difficult to ignore the corresponding EMI issues for optoelectronic systems.

Electromagnetic shielding is an effective approach to isolate strong electromagnetic radiation from sensitive devices[8]. Electromagnetic shielding materials applied in optoelectronic systems are required to provide not only high EMI shielding effectiveness (SE) and ultra-wideband shielding characteristics against intense and broadband electromagnetic radiation, but also high optical transparency to ensure sufficient optical transmission for clear visual observation or high-quality imaging. In consideration of the functional requirements for attenuating microwaves and transmitting visible light, the essence of transparent electromagnetic shielding is to provide a frequency selective filtering function capable of forming a passband for optical signals and a stopband for microwaves. Practice has proved that the requirements for electromagnetic shielding and optical transmittance can be met simultaneously through the rational structure design of various transparent conductive materials[9-13]. To date, considerable studies have attempted to develop transparent EMI shielding materials based on metals,[10] polymers[9,14], nanomaterials[15,16], and composites[8,11,17], which can be categorized into continuous conductive thin films and apertured type conductive materials based on their conductive structures.

However, decoupling the restrictive performance factors of optical transmittance and EMI shielding effectiveness is proving difficult. For continuous conductive thin films, a higher electrical conductivity improves the EMI SE, while reduces the optical transparency. For example, continuous ITO and ultrathin metallic films (e.g., thickness < 20 nm) can achieve transparency and be employed as transparent electromagnetic shields[10,18-20], the light transmission of ultrathin metal films depends heavily on their thickness due to light absorption or reflection. It follows that achieving superior shielding performance at such thickness is difficult since efficient shielding requires much thicker material than the skin depth. Conductive polymers, as a promising flexible and stretchable electromagnetic shielding materials[9,14], its transparency also suffers from the exponential decay of light transmission with increasing thickness. Due to the limitation of intrinsic conductivity, developing next-generation polymer-based EMI shielding materials relies on rational structural design and new preparation methods[21]. The emerging nanomaterials graphene, presents high transparency of 97.7%, while its shielding is low due to its high sheet resistance[11,15,22-24], and its combination with other shielding materials are proved practical[13,23,25].

With analysis of above continuous conductive films, the greater the thickness relative to the wavelength, the less EM waves are transmitted. In the other hand, achieving high EMI shielding performance requires lower sheet resistance, i.e., higher carrier concentration or mobility, which enhances the interaction of photons with electrons in the material and decreases light transmission under high photon absorption. Consequently, a trade-off has to be made between the optical transmittance and electrical conductivity in developing high-performance transparent conductive materials, and their applications in EMI shielding are constrained by the inability to concurrently combine high light transmission and high shielding performance.

The restriction between light transmittance and EMI shielding also exists in apertured type conductive materials such as metallic nanowires (NWs) and carbon nanotubes (CNTs) [26-31]. For instance, the Ag NW-based film can achieve a shielding effectiveness of 20.7 dB while maintaining a light transmittance of 92%[26]. Although the shielding performance can be further enhanced by increasing the loading amount of Ag NWs, its light transmission will inevitably decrease. Likewise, CNT-based films also require an adjustment to the content of the material to balance light transmission and electrical conductivity. Another aperture type of transparent EMI film is metallic mesh, ranging from simple square grids to complex cracked meshes, as well as grids with stacked structures[32-36]. However, conventional mesh-based transparent EMI shielding materials still suffer from the inability of integrating outstanding EMI SE with high optical transmittance. The light transmittance of metallic mesh is linearly related to the aperture area percentages of the whole film. An increase in aperture area results in a higher sheet resistance, which consequently translates into a reduction in shielding effectiveness. Besides, as the operating frequency increases, the shielding effectiveness of the metal grid shielding material diminishes dramatically as a result of filtering characteristics of meshes[37,38], which remains a barrier to the realization of efficient electromagnetic shielding over a broad frequency band.

The difficulty in designing optoelectronic windows with high transmittance and efficient EMI SE across a broad frequency range lies in: (i) separating the correlation parameters between optical transmission and EMI shielding via a decoupling design and overcoming the mutual constraints between transmittance and shielding efficiency, and (ii) constructing a rational structure to reduce the dependence of shielding efficiency and wavelength in order to obtain a high shielding efficiency in ultra-wide frequency spectrum. So far, to the best of our knowledge, obtaining electromagnetic functional materials integrated high visible light transmittance with superior EMI shielding properties in an ultra-wide frequency spectrum remains a major challenge.

In this work, we propose a design strategy of shorted micro-waveguides (SMWs) array aimed at decoupling the optical transmittance and EMI SE for electromagnetic shielding windows. In the designed transparent EMI shielding materials, the shielding layer of microstructures has unique features of microscale aperture, thin pore-wall, and large aspect ratio, which enables superior EMI SE while maintain high optical transmittance. SMWs array obtained by assembling apertured micro-waveguides shielding layer and continuous transparent conductive film further enhances shielding effectiveness and extends the effective shielding frequency spectrum. Through the optimal decoupling design, the mutual suppression indicators of light transmittance and EMI SE, as well as wideband shielding features have been incorporated into the shorted micro-waveguides array, exhibiting a promising design strategy on electromagnetic shielding for optical channels in imaging and detecting applications.

## Results

### Design Strategies of SMWs array

According to the waveguide theory, cut-off waveguides have characteristics that can be described by their cut-off frequencies. Electromagnetic waves with a frequency higher than the cut-off frequency can propagate in the waveguide, otherwise the waves are cut-off and attenuated[39]. As previously stated, the essence of transparent electromagnetic shielding is to provide a passband for light and a stopband for microwaves, which can be realized by waveguides with micron-sized aperture. The micro-waveguide for transparent EMI shielding can be designed with a cut-off frequency much higher than that of the waves to be shielded while lower than optical frequencies, allowing a substantial attenuation of the EM wave as it passes through it while ensuring the effective transmission of light.

Here, microstructures with large apertures and thin pore-wall are designed in the first place. Considering the obscuration rate and stray light uniformity[40], the efficient geometry of honeycomb was adopted. Each unit of microstructures in this case can be considered as a micro-scale cut-off waveguide owing to the large aspect ratio feature as shown in Fig. 1a. The cut-off effect of micro-waveguides can be understood as follows[41]: EM waves propagate in a zigzag pattern along the waveguide tend to be perpendicular to the wall of the waveguide under waveguide boundary conditions. The lower the frequency, the larger the angle of reflection. When the frequency reaches the cut-off frequency, EM waves bounces up and down the wall and then stops propagating forward with all energy dissipating in the conductor.

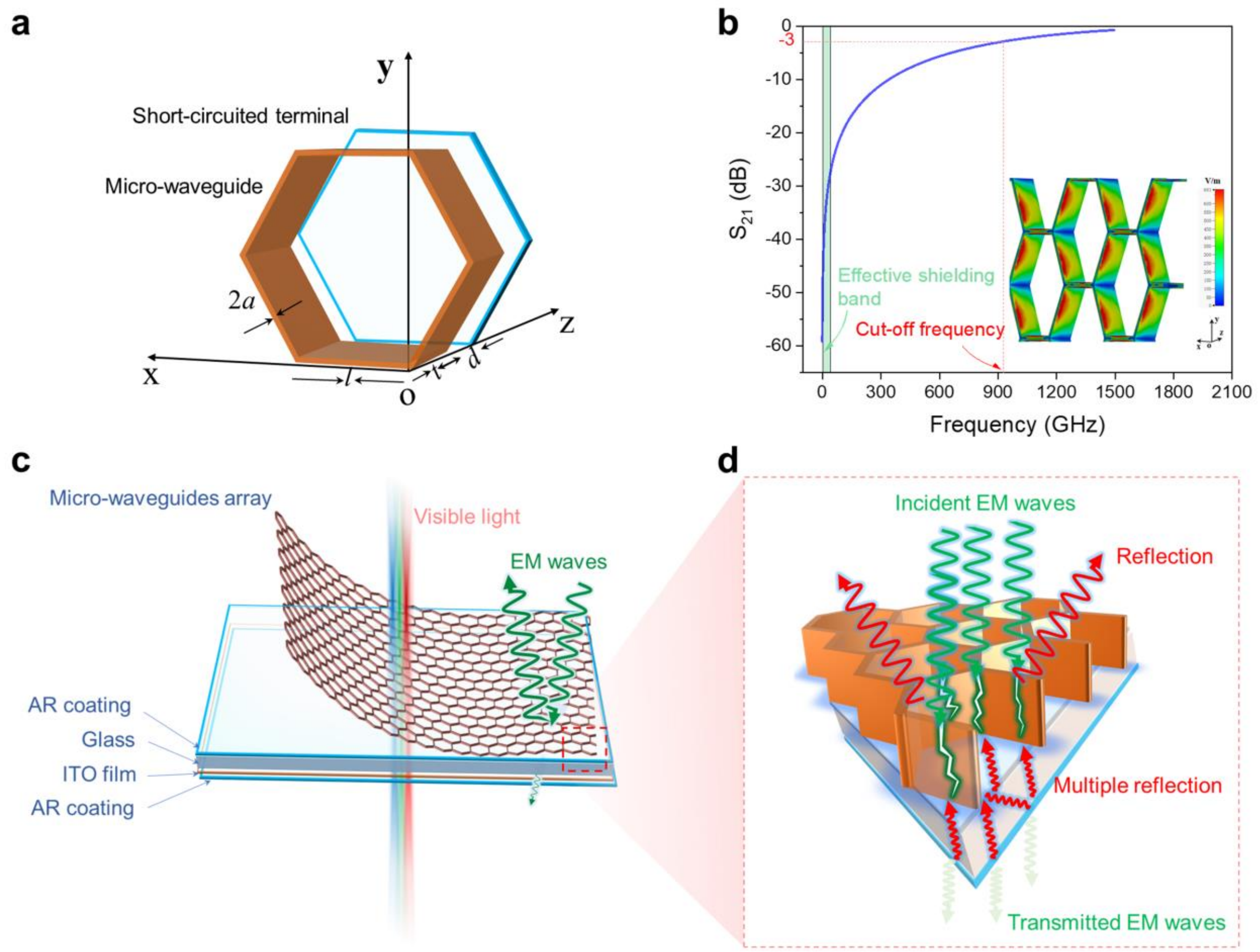

**Fig. 1 Design concept of SMW array. a** The unit structure of shorted micro-waveguides array. **b** The simulated transmission characteristic of micro-waveguide, and the inset is the electric field distribution of EM waves at 10 GHz. **c** Schematic diagram structure of shorted micro-waveguides array for achieving high optical transparency and superior electromagnetic shielding. **d** Principle diagram of EMI shielding mechanism of shorted micro-waveguides array

It is feasible to establish a cut-off filtering property for microwaves and efficiently shield against microwaves while providing a passband for light waves and keeping a high optical transmission rate by engineering the structural parameters of the metallic micro-waveguides array. According to the theorical analysis, micro-waveguides with an aperture side length of 80 μm and pore-wall thickness of 4.3 μm have a cut-off frequency up to 0.924 THz, which is much higher than those of the microwaves to be shielded. (The estimated calculation of the cut-off frequency can be seen in Supplementary Note 1 for details). As illustrated in Fig. 1b, effective electromagnetic shielding can be achieved in the cut-off region where the operating frequency band is far from the cut-off frequency. The inset of Fig. 1b shows the simulation result of field distribution in the hexagonal micro-waveguide, revealing the effective attenuation of EM waves along the short-circuited waveguides. Simultaneously, the large aperture and thin pore-wall provide a high obscuration ratio and ensures a high rate of light transmission. Consequently, the cut-off effect of micro-waveguide array yields enhanced electromagnetic attenuation without sacrificing optical transmission, allowing for a decoupling of light transmission and EMI shielding property.

EM attenuation of the micro-waveguides decreases in higher frequency bands (Fig. 1b), revealing the difficulty to get high SE over broadband frequencies. To address this problem, a continuous transparent conductive layer (ITO) is incorporated into the metallic apertured microstructures to constitute a shorted micro-waveguides array, in which the ITO film serves as the short-circuited terminal (Fig. 1a). For a perfect shorted waveguide in electromagnetic theory, its short-circuited load allows all electromagnetic waves to be reflected back. ITO films with low sheet resistance can provide efficient EMI shielding as well as high transparency, which are suitable as proper short-circuited terminals for SMWs array. The selection of short-circuited film for SMWs array can be found in Section 2 of Supplementary Information. The shorted micro-waveguides configuration can generate effective EM reflection and attenuation, and compensates for the shielding attenuation of the separate micro-waveguide in higher frequencies through multiple reflection excited between the two conducting layers. The synergistic attenuating effect of the shorted micro-waveguides is expected to produce an enhanced and balanced shielding over an ultra-wide frequency band. Additionally, to further reduce the optical loss of the shorted micro-waveguide array, anti-reflective (AR) layers for visible light could be introduced, which serve the purpose of regulating the optical transmission characteristics of the combined shielding structure and increasing the light transmission rate without lowering the shielding efficiency.

Motivated by the design concept above, as in Fig. 1c, the highly transparent EMI shielding material based on SMWs array consists primarily of a combined shielding structure of hexagonal metallic micro-waveguides and ITO continuous film, separated by thin quartz glass, and the ITO glass is covered with AR coatings. Specific AR films are designed as shown in Fig. S1 (Supplementary Information). Fig. 1d schematically illustrates the electromagnetic shielding mechanism of the shorted micro-waveguides. When EM waves incident on the shorted micro-waveguides array, a reflection occurs on the surface of the metallic apertures due to impedance mismatching. After further attenuating in the cavity of the micro apertures, multiple reflections will be established in the shorted micro-waveguides array between the apertured layer and the continuous conductive layer, contributing to the enhancement of EMI shielding. It follows that the synergy attenuating effect of the shorted micro-waveguide configuration endows the material superior EMI shielding properties over an ultra-wide frequency spectrum.

**Performance of SMWs array**

The shielding effectiveness quantifies the degree to which electromagnetic fields or electromagnetic waves are attenuated by shielding materials. It is defined by the logarithmic ratio of the incident power to the transmitted power[13,42], and the larger the SE value, the smaller the amount of electromagnetic radiation passes through the material. Generally, the EM shielding effect can be produced by electromagnetic waves being reflected at the surface of shielding materials, being absorbed, or by multiple reflection occurring within the shielding material itself. The total EMI shielding effectiveness of the transparent EMI shielding material is the sum of EM reflection, absorption, and multiple internal reflection, and can be obtained by transmission coefficient ($T$) or the scattering parameters[42]:

$$\mathrm{SE} = 10\log\left(\frac{1}{T}\right) = 10\log\left(\frac{1}{|S_{21}|^2}\right) \tag{1}$$

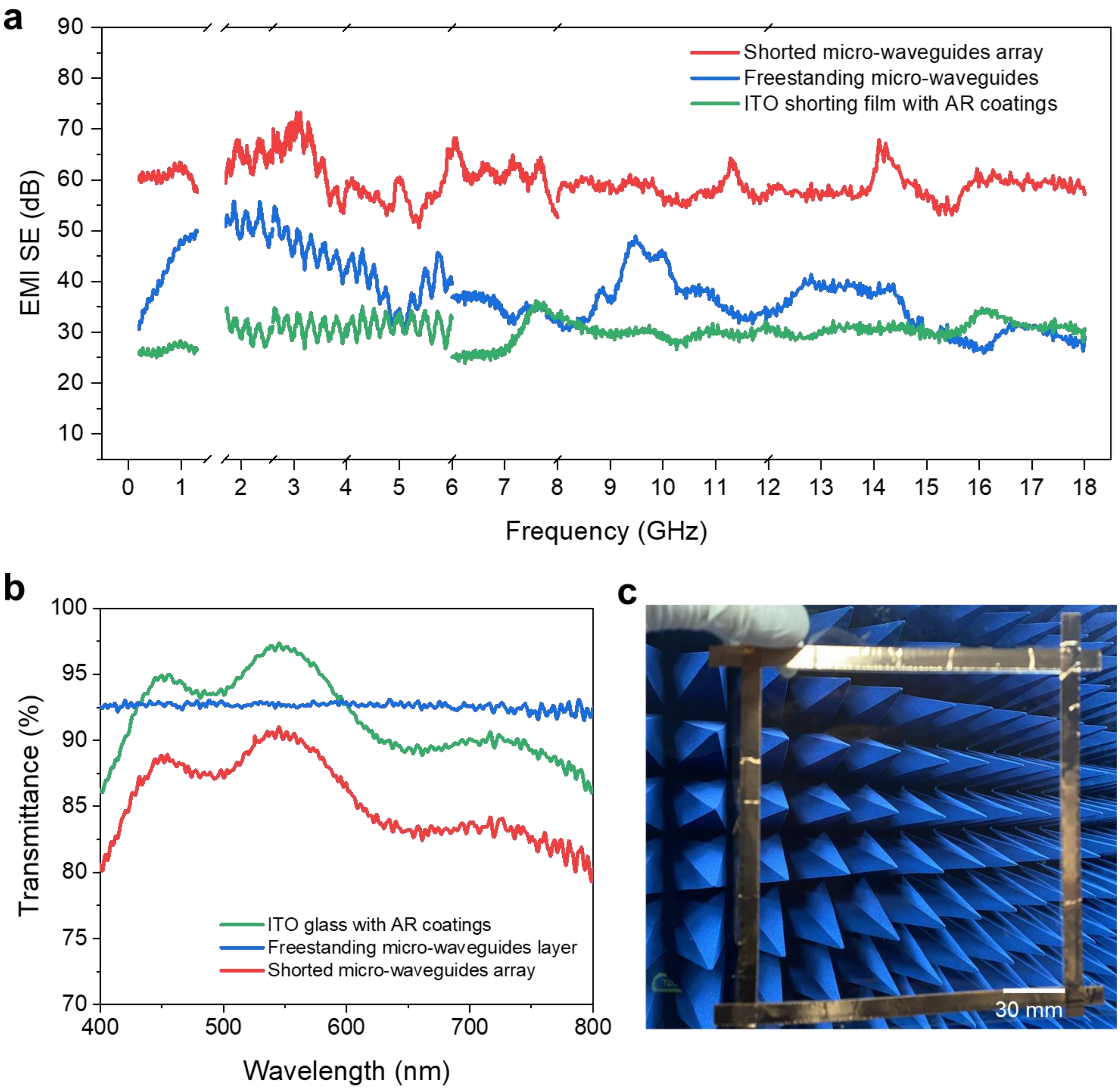

**Fig. 2 Performance of SMW array**. **a** EMI shielding performance of the freestanding micro-waveguides layer ($l = 80$ μm, $2a = 4.3$ μm, $t = 14$ μm), ITO shorting film with AR coatings, and the shorted micro-waveguides array. **b** Optical transmittance of the shorted waveguide array and its components in visible range. **c** Visible observation through the transparent EMI shielding material based on SMWs array

Figure 2a presents the EMI shielding performance of the designed transparent shield based on the array of SMWs. For comparison, EMI SE values of freestanding micro-waveguides layer and ITO shorting film are plotted as well. The freestanding micro-waveguides ($l$ = 80 μm, 2$a$ = 4.3 μm) has an average SE of ~40.7 dB in a broad frequency range. Especially in the frequency range from 1.7 GHz to 4 GHz, more efficient EMI shielding with exceptional SE exceeding 40 dB is obtained owing to the cut-off effect of micro-waveguides. It should be noted that, the freestanding micro-waveguide layer demonstrates a downward trend in EMI shielding effectiveness as the frequency goes higher, exhibiting an over 30-dB reduction in SE from 1.7 GHz to 18 GHz (59.4 dB to 29.1 dB). Likewise, a steep decline in EMI SE is also observed with the frequency decreasing to megahertz spectrum.

It is associated with two factors of the frequency dependence on the sheet resistance and the perforated structure of the micro-waveguide that leads to the reduction in EMI SE with increasing frequency. The frequency dependence on sheet resistance of micro-waveguide layer can be describe by the following equation[34,43]:

$$R = \frac{1}{\sigma\delta(1 - e^{-t/\delta})}\frac{l}{2a} \tag{2}$$

Where $\delta$ is the skin depth and $\delta = \sqrt{1 / \pi f \mu \sigma}$, $\mu$ and $\sigma$ are the permeability and conductivity of the micro-waveguide layer. The active part of their sheet resistance increases with increasing frequency in proportion to the square root of the frequency[43]. The increased sheet resistance of micro-waveguides layer partly influences on the attenuating EMI shielding behavior. Additionally, the cut-off effect of micro-waveguides also contributes to the degradation of EMI SE at higher frequencies. Besides, apertures in metallic micro-waveguides cause radiation leakage due to the diffraction relationship between wavelength and aperture size. According to the study carried out by Lee et.al., it is not the number of waveguides that determines the shielding effectiveness of the hexagonal micro-waveguide array, but rather the geometrical parameters of the unit waveguide. And the shielding effectiveness of the micro-waveguides array can be described by the following equation[44]:

$$\mathrm{SE} = 17.5\frac{t}{l}\sqrt{1-\left(\frac{lf}{96659}\right)^2} - 20\log\frac{2kl}{\pi}\cos\phi \tag{3}$$

where $k$ is wavenumber, $l$ is a transverse dimension of waveguide, and $\phi$ is the angle of incident waves. It indicates that the shielding effectiveness of micro-waveguides gets weaker with the increasing frequencies. Not surprisingly, strong evidence of shielding performance degradation of the freestanding micro-waveguides was found as frequency goes higher, which is consistent with the reported shielding properties of apertured type shielding materials[36-38,45]. At frequencies of the megahertz spectrum, the wavelength of the incident electromagnetic wave is much larger than the section size of the micro-waveguide. As the frequency decreases, the influence of the cut-off waveguide effect on the shielding characteristics is gradually weakened. And the micro-waveguide layer behaves as a thick continuous conductive film when EM waves incident. At the low frequency limit of the measured frequency band (200 MHz), the calculated SE is ~29 dB (according to the equation S4 of Supplementary Note 2), which is close to the measured result (30.7 dB). It confirms that the cut-off effect of the microwave guide has little contribution to the shielding performance at megahertz frequency spectrum.

To achieve both higher level of light transmission and EMI shielding performance in broadband frequency spectrum, we fabricated shorted micro-waveguides array by assembling the metallic micro-waveguides and ITO film. What is striking about the results is that the EMI SE of the SMWs array is measured to be 62.2 dB on average in a wide frequency range (0.2–1.3 GHz & 1.7–18 GHz) while keeping an outstanding optical transmittance of 90.4% at 550 nm. More importantly, there was no significant shielding degradation as the frequency changes, demonstrating the indiscriminate shielding property over a broad frequency range. The remarkable EMI shielding performance of SMWs array originates from the enhanced attenuation by the waveguide effect and multiple internal reflections from the layered structures.

In comparison to the pristine micro-waveguide layer, the SMW-based structures are more effective at attenuating microwaves owing to the introduced ITO film. As shown in Fig. 2a, the ITO shorting film exhibits a stable EMI SE of ~30 dB over the entire frequency. The ITO continuous film with high electrical conductivity functions as the short-circuited terminal for the micro-waveguides, which can reflect EM waves back to the micro-waveguides and arose additional reflection on the interior interfaces for further attenuation (Fig. 1d). The incident microwaves can be adequately shielded in the low frequency region due to the reflection from the surface of the micro-waveguides and the further attenuation of the apertured microstructure by waveguide effects. The microwave energy contained within the micro-waveguide is converted to heat and dissipated. Microwave energy transmitted through apertures stimulates multiple reflections between the micro-waveguides and the ITO shorting film, hence increasing the shielding effectiveness. The shielding enhancement effect is restricted in lower frequencies due to the dominant shielding property of the micro-waveguide. In the high frequency range, the shielding efficacy of freestanding micro-waveguides drops, and more microwaves enter the layered structure, resulting in constructive interference and effective multiple reflection, thereby exhibiting extremely high overall EMI SE as well. As a result of the synergistic effect of the assembled shielding materials in the SMWs array, the EMI attenuating effect of the freestanding micro-waveguide in the higher frequency band is enhanced and strong and balanced microwave shielding performance in the wide frequency band is achieved.

Optical transmittance is a fundamental and essential performance indicator of transparent electromagnetic shielding material to ensure that the shielding material does not impair the signal transmission of the optoelectronic system itself. The optical transmittance curves of freestanding micro-waveguides, short-circuited terminal (glass with ITO and AR coatings), and the SMW array are shown in Fig. 2b. By depositing properly designed AR layers onto the ITO glass (see of Supplementary Fig. S1 for details), the optimized ITO shorting film exhibits an average transmittance of 91.0% in the wavelength range of visible light (400–800 nm) and an exceptional transmittance of 96.1% at 550 nm. The optical AR layers modulate the transmission properties of visible light and further reduced the optical loss, especially around 550 nm, thereby contributing to the highly transparency of ITO shorting films. Apertured materials have total transmittance equal to each diffracted order's transmittance, and can be expressed as the obscuration ratio[38]. That is, the total optical transmittance of the freestanding micro-waveguides layer is approximately equal to the ratio of the area covered by non-metallic material to the entire area. The large opening fraction of the micro-waveguide layer (large side length of 80 μm and fine pore-wall thickness of 4.3 μm) allows more light transmitting through the material to achieve a high transmittance. Consequently, the average transmittance of the freestanding micro-waveguides was measured to be 93.1%. Remarkably, a high transmittance of 90.43% at 550 nm is maintained by the SMW array owing to the high transparency of its two functional components. Figure 2c shows the digital photograph of the fabricated shorted micro-waveguide array. High optical transmission ensures good visual effect, allowing clear observation of the scene behind the shielding materials.

It is worth noting that the insulation distance ($d$) of SMW is a contributing factor in determining the constructive interference degree of multiple interspace reflection, and the intensity is greatest when $d$ equals to the quarter-wave.[46] We evaluated the shielding enhancement effect of different insulation distances on the performance of the shorted micro-waveguides array in Fig. 3. The shielding performance has been improved by increasing insulation distance, however, the degree of enhancement varies between the high and low frequency bands. In the lower frequency band (1.7–4 GHz, Fig. 3a), the shielding effectiveness increases by 14.3% with $d$ changing from 0.2 mm to 1 mm, exhibiting a limited enhancement effect. Satisfyingly, the average shielding effectiveness can be effectively enhanced by 18.6 dB under the same setting in X- and Ku-band (Fig. 3b), thereby providing a comparable shielding property in higher frequencies. In addition, the trend of improvement in SE becomes slowing down with the further increase of $d$, which is caused by the fact that equivalent reflectivity of dielectric substrates increases gradually as the equivalent optical thickness of the substrate increases from zero to quarter-wave[35]. The result verifies the critical influence of the insulation thickness between the metallic micro-waveguides and the ITO continuous conductive layer on the shielding performance, and obtained the shielding performance that can be comparable in the high and low frequency bands under the 1-mm thickness insulation condition.

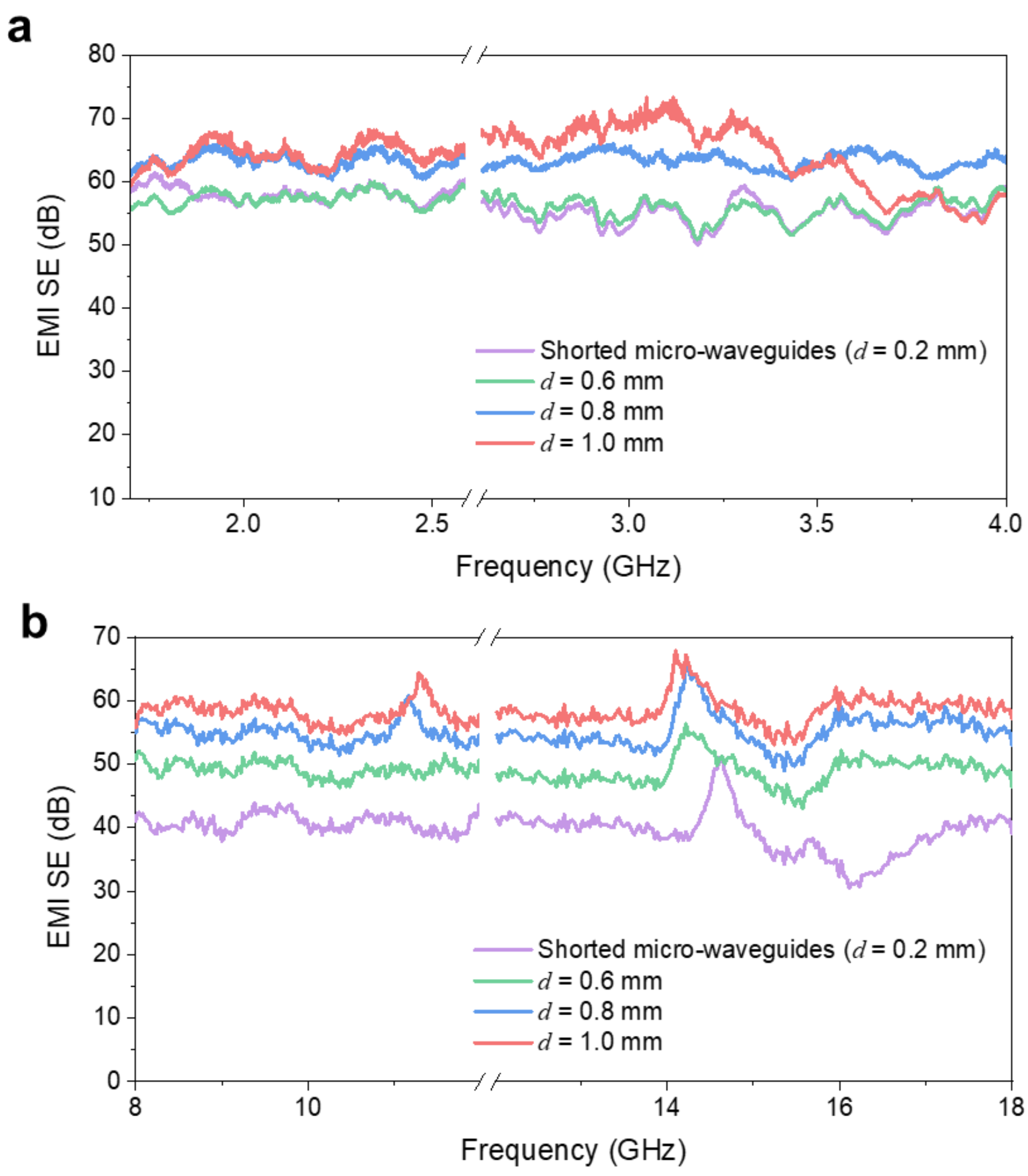

**Fig. 3 Effect of insulation distance (*d*) on EMI SE.** EMI shielding performance of SMW array with varying insulation distance (*d*) in varying frequency range. **a** 1.7–4 GHz. **b** 8–18 GHz

In addition, we investigated the performance of SMWs arrays with different structural parameters ($l$=60 μm, 65 μm, 75 μm, and $t$=14) of micro-waveguides, as shown in Fig. 4a-f.

One can see that the shielding effectiveness of SMWs slightly increases with smaller micro-waveguide apertures but is accompanied by a slight decrease in transmittance (Optical transmittances of the samples can be seen in Supplementary Fig. S2). The specific performance parameters are summarized in Table 1**.** The enhanced shielding performance is caused by the effective electromagnetic shielding band moving away from the cut-off frequency as the side length of the micro-waveguide aperture decreases. It demonstrates that the proposed shorted micro-waveguides array exhibit excellent overall performance, i.e. both high transparency and strong broadband EMI shielding. On the other hand, variatioins in structural parameters in this range produce insignificant changes in performance, providing a high degree of tolerance for material design and implementation.

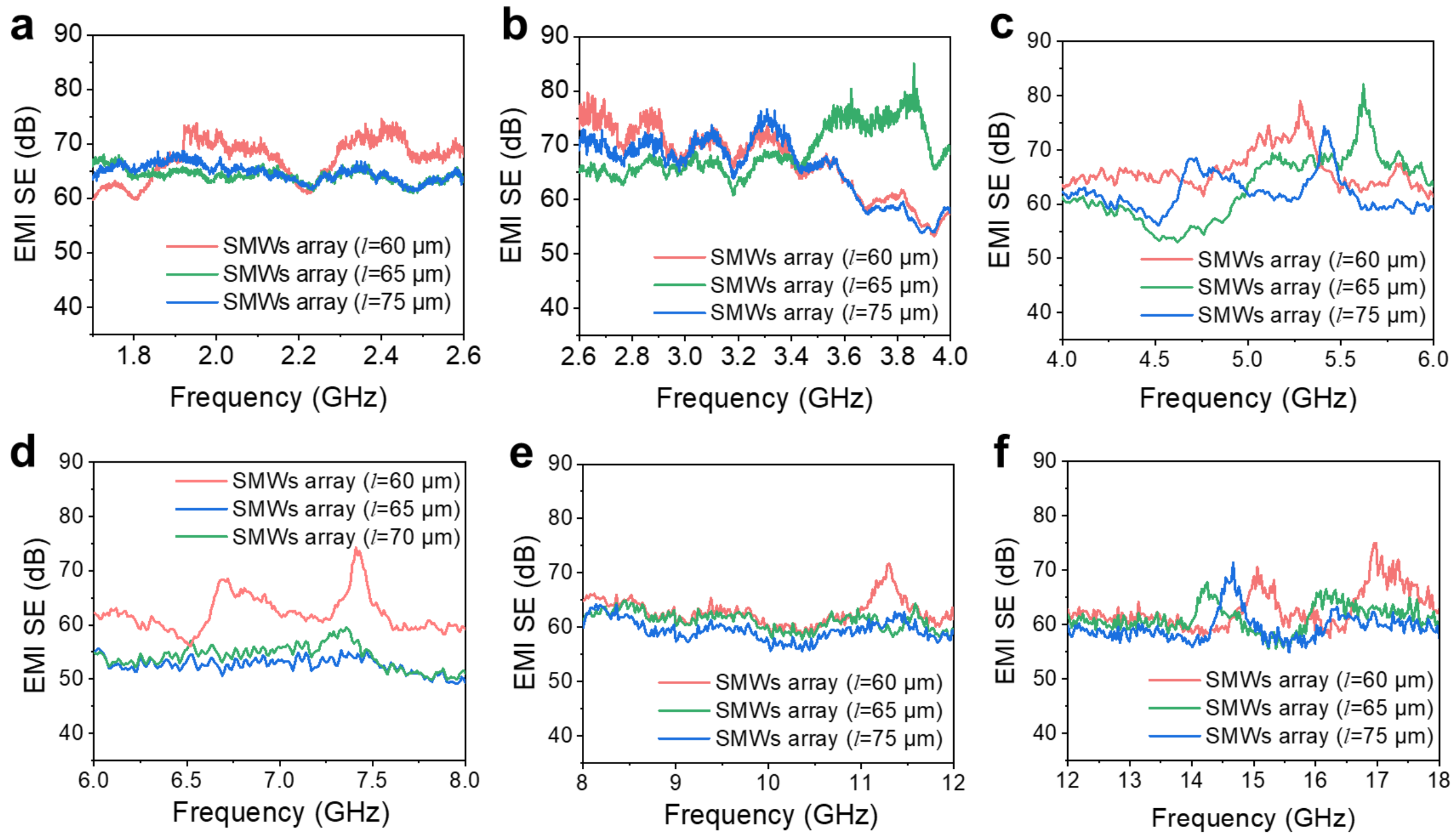

**Fig. 4 EMI SE comparison of SMW arrays (*d*=1 mm).** EMI SE of SMW arrays with varying structural parameters at different frequency band. **a** 1.7–2.6 GHz. **b** 2.6–4 GHz. **c** 4–6 GHz. **d** 6–8 GHz. **e** 8–12 GHz. **f** 12–18 GHz

**Table 1. Summary of performance merits of SMWs arrays with varying geometric dimensions**

| SMWs arrays[a] | EMI SE[b] (dB) | Transmittance[c] (%) | Frequency range (GHz) |
|---|---|---|---|
| SMWs (*l*=60 μm) | 65.9 | 87.7 | 1.7–18 |
| SMWs (*l*=65 μm) | 64.3 | 88.8 | 1.7–18 |
| SMWs (*l*=75 μm) | 63.0 | 89.5 | 1.7–18 |
| SMWs (*l*=80 μm) | 62.2 | 90.4 | 0.2–1.3, 1.7–18 |

[a)] The insulation distances of SMWs arrays were 1 mm.
[b)] The values were average shielding effectiveness obtained in the measured frequency range.
[c)] Optical transmittances were obtained at 550 nm.

To highlight the merits of this work, we have conducted a comprehensive literature review and compared the optical transmittance and EMI SE of the major transparent electromagnetic shielding materials currently available, including ultrathin metal films[18,19,47], metallic mesh films[32,34,36,37,40,48,49], and various transparent conductive composite films based on graphene[13,50], transparent conductive polymers[9,14], carbon nanotubes[51], and silver nanowires[12,26,30,52-55]. Although EMI SE exceeding 20 dB in a wide frequency band can meet commercial application standards[56], effective EMI shielding for optoelectronic system needs strengthened shielding performance. As depicted in Fig. 5, while some of the state-of-the-art works achieve extremely high transmittance (more than 90%), the associated shielding performance is limited, typically below 40 dB, making it hard to meet the requirements for high-level shielding application scenarios. By comparison, our study reveals the capacity to achieve exceptional shielding performance (average SE of 62.2 dB) while maintaining highly optical transparency, providing an outstanding comprehensive performance, and meeting the demand for EMI shielding of optoelectronic equipment. It should be noted that the proposed SMWs array outperforms previously reported highly transparent EMI shielding materials with transmittance higher than 90% in shielding effectiveness and shielding frequency band. More specific performance comparison of these transparent EMI shielding materials can be found in Supplementary Table S1.

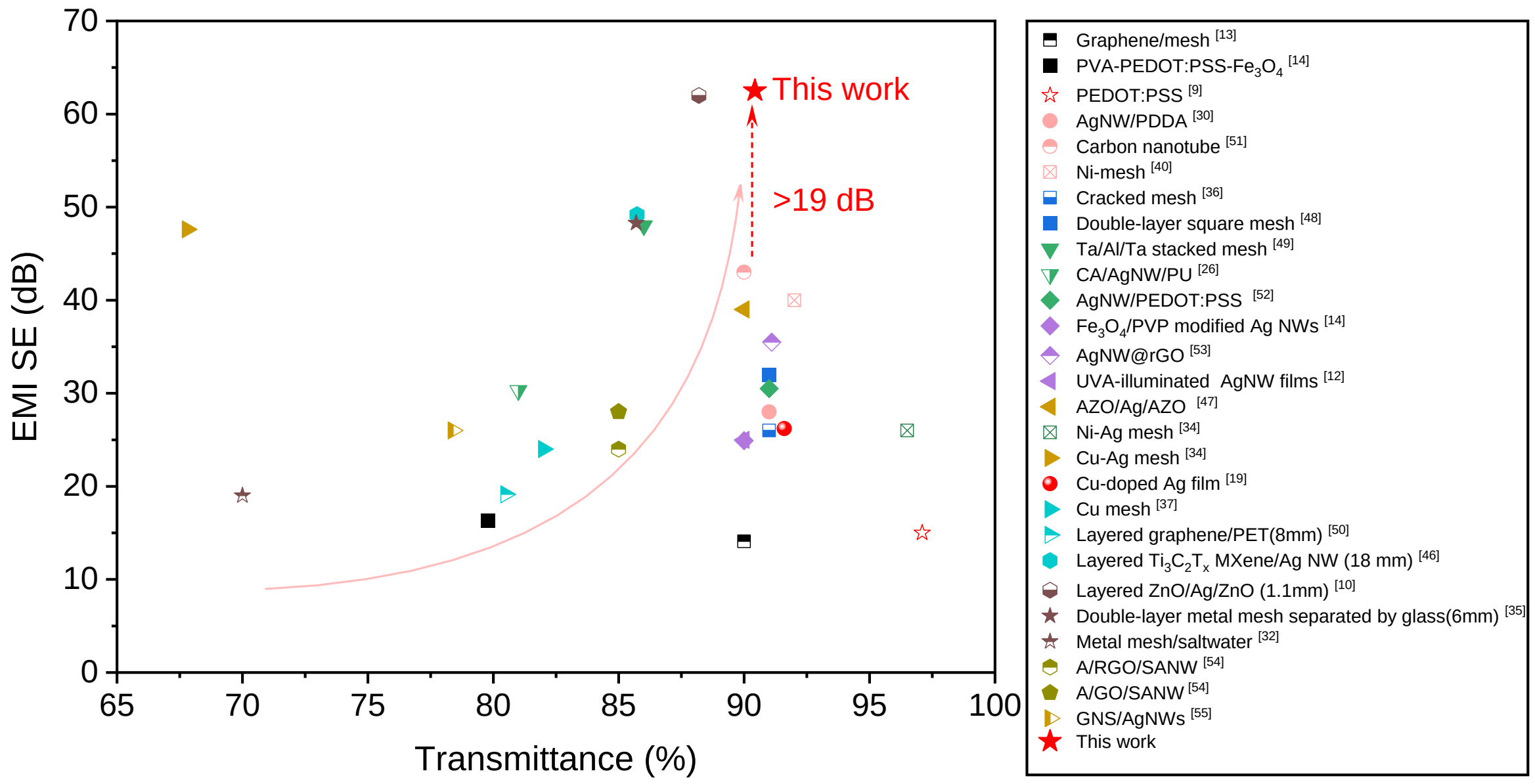

**Fig. 5 Comparison of average EMI SE and optical transmittance for our transparent EMI shielding material with previously reported results.** Our work can outperform the prior transparent EMI shielding materials at the corresponding transparency

**Comparative study on micro-waveguides array with varying structural parameters**

Due to the limit of existing lithography process, preparing apertured microstructures with a high depth-to-width ratio is challenging[57]. We adopted an additive manufacturing process to fabricate the micro-waveguide array, as described in Fig. 6a. The specific fabrication of freestanding micro-waveguides and SMWs array is given in the Experimental Section. After being peeled from the PET substrate, the freestanding micro-waveguides were achieved. The micrograph of apertures under different magnification are shown in Fig. S3 (Supplementary Information).

Optimizing the geometric parameters of the micro-waveguide is crucial for the construction of high-performance shorted micro-waveguide array. For a hexagonal micro-waveguide, the side length is a primary indicator of its aperture, which determines the conductivity and shielding ability of the micro-waveguides. Meanwhile, the variation of the aperture also directly affects the light transmission rate. In Fig. 6b, c, the shielding characteristics of four samples with different side length of 60 μm, 65 μm, 75 μm, and 85 μm are compared to investigate the specific influence of different periods on the shielding performance of the metallic apertured materials (waveguide length = 14 μm). As the side length increases from 60 μm to 85 μm, there is a 12.5-dB reduction and a 6.5-dB reduction in shielding effectiveness in the frequency ranges of 1.7–2.6 GHz and 8-12 GHz, respectively. Correspondingly, the visible light transmittance increased from 90.8% to 93.2% (at 550 nm, see Fig. 6d).

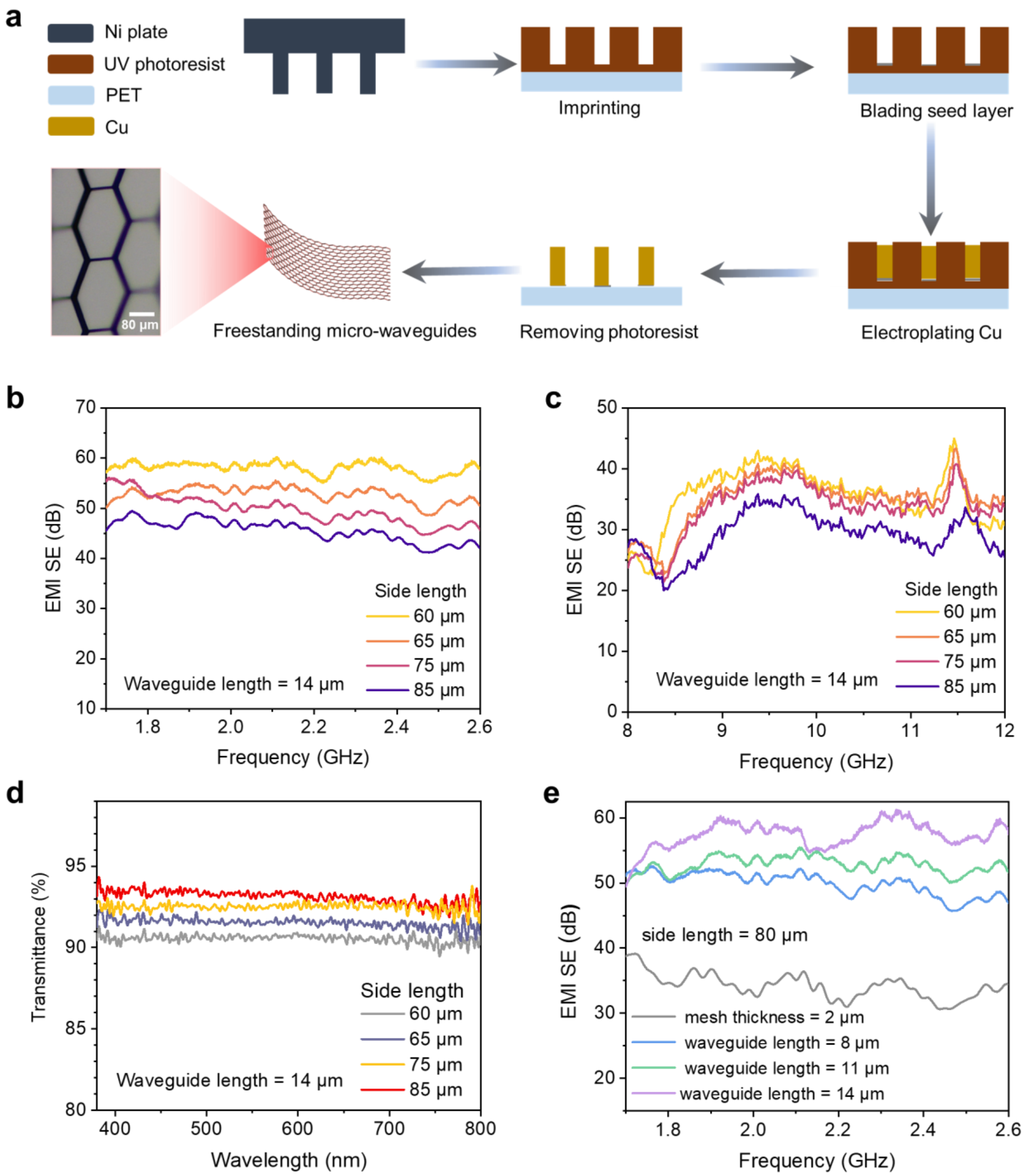

**Fig. 6 Comparative study on micro-waveguides array with varying structural parameters.** **a** Fabrication process of metallic hexagonal micro-waveguides using additive manufacturing method. The freestanding apertured micro-structures with high aspect ratios obtained by peeling off from the PET substrate. **b** EMI shielding performance of separate micro-waveguide array ($t$ = 14 μm) with varying side lengths in the frequency range of 1.7–2.6 GHz. **c** 8–12 GHz.

**d** Optical transmittance of micro-waveguides with varying side length in visible range ($t$ = 14 μm). **e** EMI SE of separate micro-waveguides with different thickness ($l$ = 80 μm)

The proposed micro-waveguide shielding layer is distinguished by its high-aspect-ratio feature in comparison to conventional metallic meshes. For micro-waveguides array with limited thickness and large aperture, the shielding efficiency can be improved by increasing the waveguide length. Longer pore walls can enhance the cut-off effect of micro-waveguides and consume the EM wave to the greatest extent. Here, we studied the shielding performance of the freestanding micro-waveguide with different waveguide length to verify the enhanced effect of waveguide effect on shielding effectiveness. Fig. 6e compares the EMI SE obtained from the prepared micro-waveguide layers with varying thicknesses (i.e., micro-waveguide length) in a lower frequency range (1.7–2.6 GHz). For these measured samples with thicknesses of 2.5 μm, 8 μm, 11 μm, and 14 μm, the side lengths and the linewidths of hexagonal apertures were fixed at 80 μm and 4.3 μm, respectively, meaning they possess the same light transmittance. The corresponding aspect ratios of the samples are 0.58, 1.86, 2.56, and 3.25, respectively. One can see that the metallic apertures with microscale waveguide features exhibit enhanced effects on EMI shielding performance as the length of the metallic micro-waveguide increases at lower frequencies. Close inspect of the Fig. shows that the sample with a waveguide length of 8 μm has an average SE value of 49.9 dB, which is 15 dB higher than that of conventional thin metal mesh film (2.5 μm in thickness). When the thickness of the aperture increases from 8 μm to 11 μm, the average SE increases by 2.9 dB over the tested frequency range and reaches 57.4 dB when the thickness further rises to 14 μm. In contrast with the micro-waveguides ($t$ = 14 μm), the thin mesh layer shows a much lower SE value of 34.1 dB, which drops nearly 12.5 dB in shielding performance.

This result demonstrates the advantages of micro-waveguide results in terms of enhanced EMI SE due to its cut-off effect, especially at lower frequencies. The observed increase in shielding effectiveness could be attributed to the fact that more EM waves are reflected back into the surrounding space due to the reduced sheet resistance of the apertured layer. Besides, the feature of large aspect ratio enhances the cut-off effect of waveguides and significantly attenuates the EM wave. However, there is no obvious waveguide effect in higher frequency band (Fig. S5, Supplementary Information). The optical transmittance appeared to be unaffected by the increase of the length of the separate micro-waveguides (Fig. S4, Supplementary Information), which is due to the unchanged ratio of the non-metal covered area to the total area. This result demonstrates that it is an effective way to enhance the shielding effectiveness by increasing the aspect ratio of separate micro-waveguide layer in lower frequencies while maintaining a high light transmittance. More importantly, it also implies that a certain degree of decoupling between optical transmittance and shielding effectiveness has been accomplished.

## Discussion

In summary, transparent EMI shielding materials featuring high visible transmittance, strong electromagnetic shielding effectiveness, and wide shielding frequency band have been developed in the proposed shorted micro-waveguides array configuration. The design strategy implemented in this scheme decouples the light transmission and EMI shielding, enabling both high optical transmittance of 90.4% and superior EMI SE of 62.2 dB over an ultra-wide frequency spectrum. The improved EMI shielding performance of apertured shielding materials with high aspect ratio at lower frequencies are also demonstrated, verifying the efficient attenuation of EM waves by micro-waveguides. The superiority of our transparent EMI shielding material based on shorted micro-waveguides array against its counterparts lies in the significantly enhanced shielding efficiency, highly optical transparency and, the ultra-wide shielding frequency spectrum. We believe that the proposed design concept of short-circuited micro-waveguides array and the structure optimization strategy would open a new avenue toward higher-level electromagnetic shielding applications in various optoelectronic systems.

## Methods

**Preparation of ITO shorting film with AR coatings**: ITO thin film (350 nm in thickness), as the short-circuited load for SMWs, was prepared by an electron beam evaporation system in the presence of oxygen with a base pressure of $2.0\times10^{-4}$ Pa, an accelerating voltage of ~7 kV, and electron beam current of 30 mA. The material used in this study was ITO pellets with nominal 99.9% purity $In_2O_3$:$SnO_2$ (95 wt% and 5 wt%, respectively). The deposition rate was ~0.10 nm $s^{-1}$ and the thickness of the deposited films was controlled using a quartz crystal thickness monitor. To reduce light reflection, silicon dioxide ($SiO_2$) layer was prepared on the ITO thin film. At the same time, tantalum pentoxide ($Ta_2O_5$) and $SiO_2$ as the high and low refractive index for the multilayer anti-reflection films were deposited on the back of substrate by electron-beam evaporator with ion-beam assisted deposition (IAD). The deposition rates were ~0.15 nm $s^{-1}$ and 0.4 nm $s^{-1}$, respectively. The substrate temperature during the deposition process was kept at 250℃. The specific film structure can be found in Supplementary Fig S1.

**Fabrication of freestanding micro-waveguides**: Firstly, the nickel master plates with designed geometrical dimensions were obtained by direct write and electroforming. Combined with the nickel master plate, the hexagonal trenches of different depths were formed on the UV curing adhesive layer after imprinting process. The optimized process parameters of UV curing energy is 20–30 mJ $cm^{-2}$, while post-curing energy is between 360 and 380 mJ $cm^{-2}$. The silver nanoparticles were filled into the embossed groove structure by blading and sintered afterwards. The following step was to electroplate Cu on top of the conductive Ag lines inside the trenches. With this method, metallic apertured microstructures with high aspect ratio greater than 3 (line width of 4.3 μm, thickness of 14 μm) were prepared. The freestanding micro-waveguides can be obtained by removing UV photoresist. To facilitate performance testing, they were fixed to the PET frame with copper tape. We obtained the SMWs array by peeling the metallic apertured micro-waveguide film from the PET substrate and transferring it to the ITO glass with AR coatings. It should be noted that the edges of the freestanding layer are particularly vulnerable to warping due to the inherent mechanical stress. All of the prepared samples are 150 mm×150 mm in size.

**Characterization and Measurements**: Optical transmittance of transparent EMI shielding materials in visible wavelength range was measured using a UV-visible-near-infrared (UV-vis-NIR) spectrophotometer (HITACHI, U-4100) in ambient condition.

To obtain the EMI shielding effectiveness over an ultrawide frequency spectrum, we utilized two methods for the EMI shielding measurement according to the measuring frequency range. The measured limits of low and high frequency were 200 MHz and18 GHz, respectively. In the frequency range of 0.2–1.3 GHz, flange coaxial method was adopted. The measurement system was connected as shown in Supplementary Fig. S6a. the input of the network analyzer is connected to one end of the flanged coaxial unit and the output is connected to the other end of the flanged coaxial unit. After calibrating the measurement system, the measuring specimen was fixed in the flanged coaxial device and the shielding effectiveness of the loaded specimen can be obtained. For the EMI shielding performance in the frequency range of 1.7–18 GHz, waveguide measuring method was employed. As shown in Fig. S6b, a vector network analyzer (VNA, AV3672B) which connected to waveguides operated in various frequency bands (1.7–2.6 GHz, 2.6–4 GHz, 4–6 GHz, 6–8 GHz, 8–12 GHz, and 12–18 GHz). The measurement system was calibrated with Transmission-Reflection-Load (TRL) technique at both ends of the waveguides. The two terminals of the vector network analyzer are connected to the waveguide via the coaxial cable, where as-prepared samples are clamped tightly between the commissure of waveguides. *S*-parameters were collected by the VNA, and total EMI SE values were calculated by equation (1).

### Acknowledgements

The authors gratefully acknowledge the funding support from the Natural Science Foundation of Hunan Province (No. 2020RC3028) and the National Natural Science Foundation of China (No. 61801490). The authors thank Prof. Haicheng Wang from GRIMAT Engineering Institute Co., Ltd. for discussions and advice about improving the manuscript.

### Authors details

[1]College of Electronic Science and Technology, National University of Defense Technology Changsha, 410073, China. [2]GRINM Guojing Advanced Materials Co., Ltd, Beijing, 100088, China

### Author contributions

X.H. conceived the ideas and designed the project. Y.L. fabricated the samples. Y.L., J.P., and W.L. performed optical and EMI shielding performance measurements. K.W. and D.Z. performed simulation analysis. P.S. designed the AR coatings. Y.L. and X.H. analysed the data and wrote the manuscript with inputs from all authors.

### Competing interests

The authors declare no competing interests.

## Additional Information

**Supplementary Information** is available for this paper.

**Correspondence** and requests for materials should be addressed to X.H.

## References

1 Jiang, C. *et al.* High-Performance and Reliable Silver Nanotube Networks for Efficient and Large-Scale Transparent Electromagnetic Interference Shielding. *ACS Appl. Mater. Interfaces* **13**, 15525-15535 (2021).

2 Chen, Y. *et al.* Ultra-thin and highly flexible cellulose nanofiber/silver nanowire conductive paper for effective electromagnetic interference shielding. *Compos. Part A Appl. Sci. Manuf.* **135**, 105960 (2020).

3 Zhang, J. H. *et al.* Energy Selective Surface With Power-Dependent Transmission Coefficient for High-Power Microwave Protection in Waveguide. *IEEE Trans. Antennas. Propag.* **67**, 2494-2502 (2019).

4 Wang, Z. *et al.* Flexible and Transparent Ferroferric Oxide-Modified Silver Nanowire Film for Efficient Electromagnetic Interference Shielding. *ACS Appl. Mater. Interfaces* **12**, 2826-2834 (2020).

5 Iqbal, A., Sambyal, P. & Koo, C. M. 2D MXenes for Electromagnetic Shielding: A Review. *Adv. Funct. Mater.* **30**, 2000883 (2020).

6 Wang, H. *et al.* Double-layer interlaced nested multi-ring array metallic mesh for high-performance transparent electromagnetic interference shielding. *Opt. Lett.* **42**, 1620-1623 (2017).

7 Schroder, A. *et al.* Analysis of High Intensity Radiated Field Coupling into Aircraft Using the Method of Moments. *IEEE Trans. Electromagn. Compat.* **56**, 113-122 (2014).

8 Lin, J., Zhang, H., Li, P., Yin, X. & Zeng, G. The Electromagnetic Shielding Effectiveness of a Low-Cost and Transparent Stainless Steel Fiber/Silicone Resin Composite. *IEEE Trans. Electromagn. Compat.* **56**, 328-334 (2014).

9 Hosseini, E., Arjmand, M., Sundararaj, U. & Karan, K. Filler-Free Conducting Polymers as a New Class of Transparent Electromagnetic Interference Shields. *ACS Appl. Mater. Interfaces* **12**, 28596-28606 (2020).

10 Yuan, C. *et al.* Record-High Transparent Electromagnetic Interference Shielding Achieved by Simultaneous Microwave Fabry-Perot Interference and Optical Antireflection. *ACS Appl. Mater. Interfaces* **12**, 26659-26669 (2020).

11 Weng, G.-M. *et al.* Layer-by-Layer Assembly of Cross-Functional Semi-transparent MXene-Carbon Nanotubes Composite Films for Next-Generation Electromagnetic Interference Shielding. *Adv. Funct. Mater.* **28**, 1803360 (2018).

12 Liang, X. *et al.* Facile and Efficient Welding of Silver Nanowires Based on UVA-Induced Nanoscale Photothermal Process for Roll-to-Roll Manufacturing of High-Performance Transparent Conducting Films. *Adv. Mater. Interfaces.* **6**, 1801635 (2019).

13 Han, Y., Liu, Y., Han, L., Lin, J. & Jin, P. High-performance hierarchical graphene/metal-mesh film for optically transparent electromagnetic interference shielding. *Carbon* **115**, 34-42 (2017).

14 Ray, B., Parmar, S., Date, K. & Datar, S. Optically transparent polymer composites: A study on the influence of filler/dopant on electromagnetic interference shielding mechanism. *J. Appl. Polym. Sci.* **138**, 50255 (2021).

15 Hong, S. K. *et al.* Electromagnetic interference shielding effectiveness of monolayer graphene. *Nanotechnology* **23**, 455704 (2012).

16 Valentini, L., Bon, S. B. & Kenny, J. M. Electrodeposited carbon nanotubes as template for the preparation of semi-transparent conductive thin films by in situ polymerization of methyl methacrylate. *Carbon* **45**, 2685-2691 (2007).

17 Ray, B., Parmar, S., Date, K., Datar, S. & Ieee. *Optically Transparent EMI Shielding Nanocomposite for X band*. 2019 URSI Asia-Pacific Radio Science Conference (AP-RASC). New Delhi, India, (2019).
18 Maniyara, R. A., Mkhitaryan, V. K., Chen, T. L., Ghosh, D. S. & Pruneri, V. An antireflection transparent conductor with ultralow optical loss (<2 %) and electrical resistance (<6 Omega sq(-1)). *Nat. Commun.* **7**, 13771 (2016).
19 Wang, H. *et al.* Highly Transparent and Broadband Electromagnetic Interference Shielding Based on Ultrathin Doped Ag and Conducting Oxides Hybrid Film Structures. *ACS Appl. Mater. Interfaces* **11**, 11782-11791 (2019).
20 Kim, Y. *et al.* Transparent and Flexible Electromagnetic Interference Shielding Film Using ITO Nanobranches by Internal Scattering. *ACS Appl. Mater. Interfaces* **13**, 61413-61421 (2021).
21 Liang, C. *et al.* Structural Design Strategies of Polymer Matrix Composites for Electromagnetic Interference Shielding: A Review. *Nanomicro Lett* **13**, 181 (2021).
22 Kang, J. *et al.* High-performance near-field electromagnetic wave attenuation in ultra-thin and transparent graphene films. *2D Mater.* **4**, 025003 (2017).
23 Ma, L. *et al.* Transparent Conducting Graphene Hybrid Films To Improve Electromagnetic Interference (EMI) Shielding Performance of Graphene. *ACS Appl. Mater. Interfaces* **9**, 34221-34229 (2017).
24 Bian, W., Tuncer, H. M., Katsounaros, A., Wu, W. & Yang, H. Microwave Absorption and Radiation from Large-area Multilayer CVD Graphene. *Carbon* **77**, 814-822 (2014).
25 Zhu, Y., Sun, Z., Yan, Z., Jin, Z. & Tour, J. M. Rational design of hybrid graphene films for high-performance transparent electrodes. *ACS Nano* **5**, 6472-6479 (2011).
26 Jia, L.-C. *et al.* Highly Efficient and Reliable Transparent Electromagnetic Interference Shielding, Film. *ACS Appl. Mater. Interfaces* **10**, 11941-11949 (2018).
27 Lin, S. *et al.* Room-temperature production of silver-nanofiber film for large-area, transparent and flexible surface electromagnetic interference shielding. *npj Flex. Electron.* **3**, 6 (2019).
28 Yang, H., Bai, S., Guo, X. & Wang, H. Robust and smooth UV-curable layer overcoated AgNW flexible transparent conductor for EMI shielding and film heater. *Appl. Surf. Sci.* **483**, 888-894 (2019).
29 Weng, C. *et al.* Buckled AgNW/MXene hybrid hierarchical sponges for high-performance electromagnetic interference shielding. *Nanoscale* **11**, 22804-22812 (2019).
30 Zhu, X. *et al.* Highly efficient and stable transparent electromagnetic interference shielding films based on silver nanowires. *Nanoscale* **12**, 14589-14597 (2020).
31 Xie, Q. *et al.* Transparent, Flexible, and Stable Polyethersulfone/Copper-Nanowires/Polyethylene Terephthalate Sandwich-Structured Films for High-Performance Electromagnetic Interference Shielding. *Adv. Eng. Mater.* **23**, 2100283 (2021).
32 Duy Tung, P. & Jung, C. W. Optically transparent and very thin structure against electromagnetic pulse (EMP) using metal mesh and saltwater for shielding windows. *Sci. Rep.* **11**, 2603 (2021).
33 Xu, X., Lin, Z., Wang, S. & Wu, S. Effect of the Rotation Angle in Multi-Ring Metallic Meshes on Shielding Effectiveness. *IEEE Microw. Wirel. Compon. Lett.* **30**, 629-632 (2020).
34 Voronin, A. S. *et al.* Cu-Ag and Ni-Ag meshes based on cracked template as efficient transparent electromagnetic shielding coating with excellent mechanical performance. *J. Mater. Sci.* **56**, 14741-14762 (2021).

35 Zhang, Y. *et al.* Double-layer metal mesh etched by femtosecond laser for high-performance electromagnetic interference shielding window. *RSC Adv.* **9**, 22282-22287 (2019).
36 Han, Y. *et al.* Crackle template based metallic mesh with highly homogeneous light transmission for high-performance transparent EMI shielding. *Sci. Rep.* **6**, 25601 (2016).
37 Han, Y. *et al.* In Situ Surface Oxidized Copper Mesh Electrodes for High-Performance Transparent Electrical Heating and Electromagnetic Interference Shielding. *Adv. Electron. Mater.* **4**, 1800156 (2018).
38 Wang, W., Bai, B., Zhou, Q., Ni, K. & Lin, H. Petal-shaped metallic mesh with high electromagnetic shielding efficiency and smoothed uniform diffraction. *Opt. Mater. Express* **8**, 3485-3493 (2018).
39 Tan, D. *et al.* Development and current situation of flexible and transparent EM shielding materials. *J. Mater. Sci.* **32**, 25603-25630 (2021).
40 Jiang, Z.-P., Huang, W., Chen, L.-S. & Liu, Y.-H. Ultrathin, lightweight, and freestanding metallic mesh for transparent electromagnetic interference shielding. *Opt. Express* **27**, 24194-24206 (2019).
41 Tan, D. *et al.* Development and current situation of flexible and transparent EM shielding materials. *J. Mater. Sci.: Mater. Electron.* **32**, 25603-25630 (2021).
42 Kim, S. *et al.* Electromagnetic Interference (EMI) Transparent Shielding of Reduced Graphene Oxide (RGO) Interleaved Structure Fabricated by Electrophoretic Deposition. *ACS Appl. Mater. Interfaces* **6**, 17647-17653 (2014).
43 Liu, Y. & Tan, J. Frequency dependent model of sheet resistance and effect analysis on shielding effectiveness of transparent conductive mesh coatings. *Progress in Electromagnetics Research* **140**, 353-368 (2013).
44 Lee, K. W., Cheong, Y. C., Hong, I. P. & Yook, J. G. Design Equation of Shielding Effectiveness of Honeycomb. Proceedings of ISAP2005, Seoul, Korea, (2005).
45 Van Viet, T. *et al.* Electromagnetic Interference Shielding by Transparent Graphene/ Nickel Mesh Films. *ACS Appl. Nano Mater.* **3**, 7474-7481 (2020).
46 Chen, W., Liu, L.-X., Zhang, H.-B. & Yu, Z.-Z. Flexible, Transparent, and Conductive Ti3C2Tx MXene-Silver Nanowire Films with Smart Acoustic Sensitivity for High-Performance Electromagnetic Interference Shielding. *ACS Nano* **14**, 16643-16653 (2020).
47 Choi, H.-J., Park, B.-J., Eom, J.-H. & Yoon, S.-G. Simultaneous realization of electromagnetic interference shielding, hydrophobic qualities, and strong antibacterial activity for transparent electronic devices. *Curr. Appl. Phys.* **16**, 1642-1648 (2016).
48 Lu, Z., Wang, H., Tan, J. & Lin, S. Microwave shielding enhancement of high-transparency, double-layer, submillimeter-period metallic mesh. *Appl. Phys. Lett.* **105**, 241904 (2014).
49 Kumar, P., Reddy, P. V., Choudhury, B., Chowdhury, P. & Barshilia, H. C. Transparent conductive Ta/Al/Ta-grid electrode for optoelectronic and electromagnetic interference shielding applications. *Thin Solid Films* **612**, 350-357 (2016).
50 Lu, Z., Ma, L., Tan, J., Wang, H. & Ding, X. Transparent multi-layer graphene/polyethylene terephthalate structures with excellent microwave absorption and electromagnetic interference shielding performance. *Nanoscale* **8**, 16684-16693 (2016).
51 Xu, H., Anlage, S. M., Hu, L. & Gruner, G. Microwave shielding of transparent and conducting single-walled carbon nanotube films. *Appl. Phys. Lett.* **90**, 183119 (2007).

52 Qin, F. *et al.* Highly Uniform and Stable Transparent Electromagnetic Interference Shielding Film Based on Silver Nanowire-PEDOT:PSS Composite for High Power Microwave Shielding. *Macromol. Mater. Eng.* **306**, 2000607 (2020).
53 Yang, Y. *et al.* Reduced Graphene Oxide Conformally Wrapped Silver Nanowire Networks for Flexible Transparent Heating and Electromagnetic Interference Shielding. *ACS Nano* **14**, 8754-8765 (2020).
54 Kim, D. G., Choi, J. H., Choi, D.-K. & Kim, S. W. Highly Bendable and Durable Transparent Electromagnetic Interference Shielding Film Prepared by Wet Sintering of Silver Nanowires. *ACS Appl. Mater. Interfaces* **10**, 29730-29740 (2018).
55 Zhang, N. *et al.* Flexible and transparent graphene/silver-nanowires composite film for high electromagnetic interference shielding effectiveness. *Sci. Bull.* **64**, 540-546 (2019).
56 Shu, J. C., Cao, W. Q. & Cao, M. S. Diverse Metal–Organic Framework Architectures for Electromagnetic Absorbers and Shielding. *Adv. Funct. Mater.* **31**, 2100470 (2021).
57 Chen, X. *et al.* Printable High‐Aspect Ratio and High‐Resolution Cu Grid Flexible Transparent Conductive Film with Figure of Merit over 80 000. *Adv. Electron. Mater.* **5**, 1800991 (2019).

Supplementary Information for

# Shorted Micro-Waveguide Array for High Optical Transparency and Superior Electromagnetic Shielding in Ultra-Wideband Frequency Spectrum

[1]College of Electronic Science and Technology, National University of Defense Technology

Changsha, 410073, China

[2]GRINM Guojing Advanced Materials Co., Ltd, Beijing, 100088, China

*To whom correspondence and requests for materials should be addressed:

huangxianjun@nudt.edu.cn; shangpeng163@163.com

## Supplementary Note 1: Cut-off frequency of hexagonal micro-waveguides

Conventional waveguides are typically designed on millimeter scale with cut-off frequencies in the microwave band. Similarly, it is possible to make an estimate for the cut-off frequency of the hexagonal micro-waveguide. Here, the cut-off frequency ($f_c$) of the hexagonal micro-waveguide can be estimated by the following equation:[1]

$$f_c = \frac{15 \times 10^9}{w}\,(\mathrm{Hz}) \tag{S1}$$

where $w$ is the diameter of the external circle of the hexagonal waveguide. For the micro-waveguide with a side length of 80 μm, the cut-off frequency is 0.937 THz, which is consistent with the simulation results (0.924 THz).

## Supplementary Note 2: Selection of short-circuited film for SMWs

For a conductive film with permeability $\mu$, permittivity $\varepsilon$, and electrical conductivity $\sigma$, The transmission coefficient of the film is[2,3]

$$t = \frac{p\,exp\,(-\gamma_t d)}{1 - q\,exp\,(-2\gamma_t d)} \tag{S2}$$

Here, the propagation constant of the electromagnetic wave at a frequency of $\omega$ is $\gamma = \sqrt{i\omega\mu(\sigma + i\omega\varepsilon)}$, the impedance of the film is $Z = \sqrt{i\omega\mu/(\sigma + i\omega\sigma)}$, the impedance of the free space is $Z_0 = \sqrt{\mu_0/\varepsilon_0} = 377\,\Omega$. $p$ and $q$ are transmission and reflection factors, respectively.

$p = \frac{4Z_0 Z_t}{(Z_0 + Z_t)^2}$, $q = \frac{(Z_0 - Z_t)^2}{(Z_0 + Z_t)^2}$, and $p + q = 1$, $d$ is the thickness of the film.

If the thickness of the film is much thinner than the skin depth, and it has a much lower impedance compared to the free space ($Z_t/Z_0 \ll 1$), equation S2 can be approximated to:

$$t = \frac{1}{1 + Z_0/(2R_s)} \tag{S3}$$

Where $R_s = 1/\sigma d$. Therefore, shielding effectiveness can be expressed as:[2]

$$\mathrm{SE} = 10\log\left(\frac{1}{t \cdot t^*}\right) = 20\log\left(1 + \frac{Z_0}{2R_s}\right) \tag{S4}$$

Based on the above expression, the shielding effectiveness of a continuous conductive film is determined by its sheet resistance and is independent of the frequency of electromagnetic wave. Thus, we can improve the EMI shielding performance by reducing the sheet resistance of the films. For the construction of shorted micro-waveguide array, transparent conductive films with high shielding effectiveness (i.e., low sheet resistance) are needed to provide efficient electromagnetic reflection as short-circuited terminal.

**Supplementary Figures**

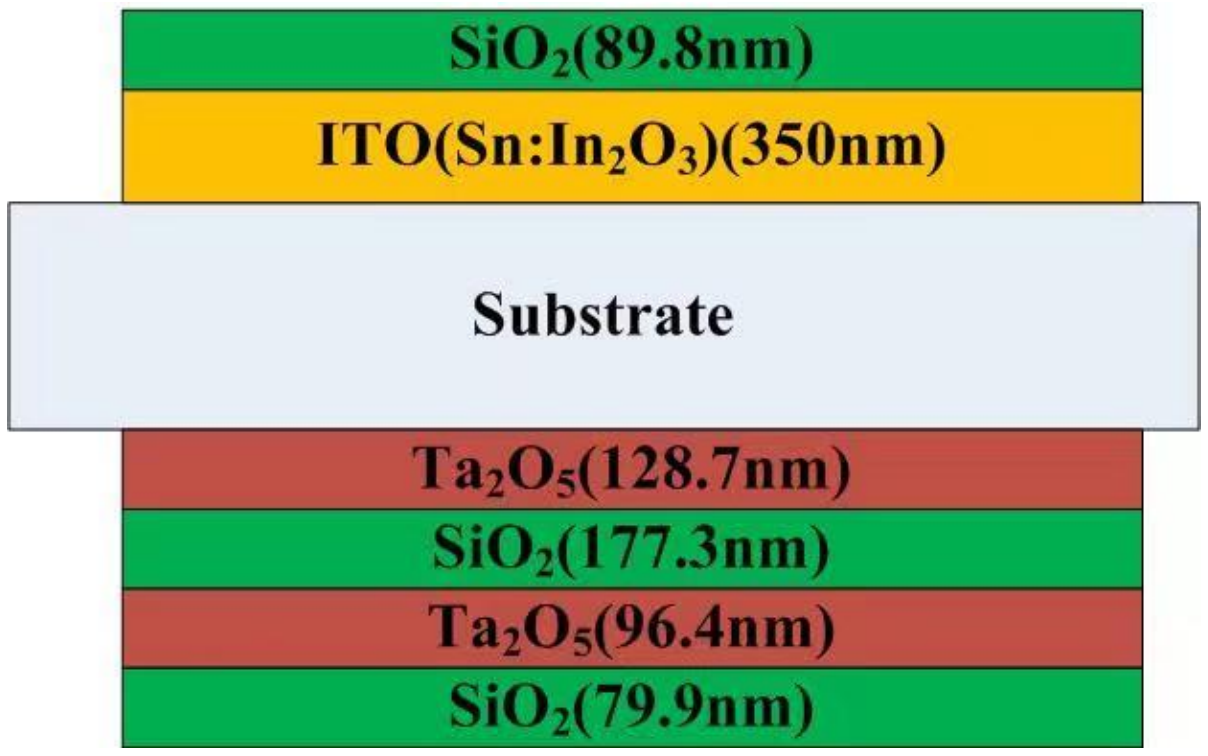

**Fig. S1 Anti-reflection design for the thin glass with ITO.** A $SiO_2$ film of 89.8 nm was deposited on the surface of ITO film. $Ta_2O_5$ and $SiO_2$ films are deposited alternately on the other side of the substate. The thicknesses of the films are 128.7 nm, 177.3 nm, 96.4 nm, and 79.9 nm, respectively.

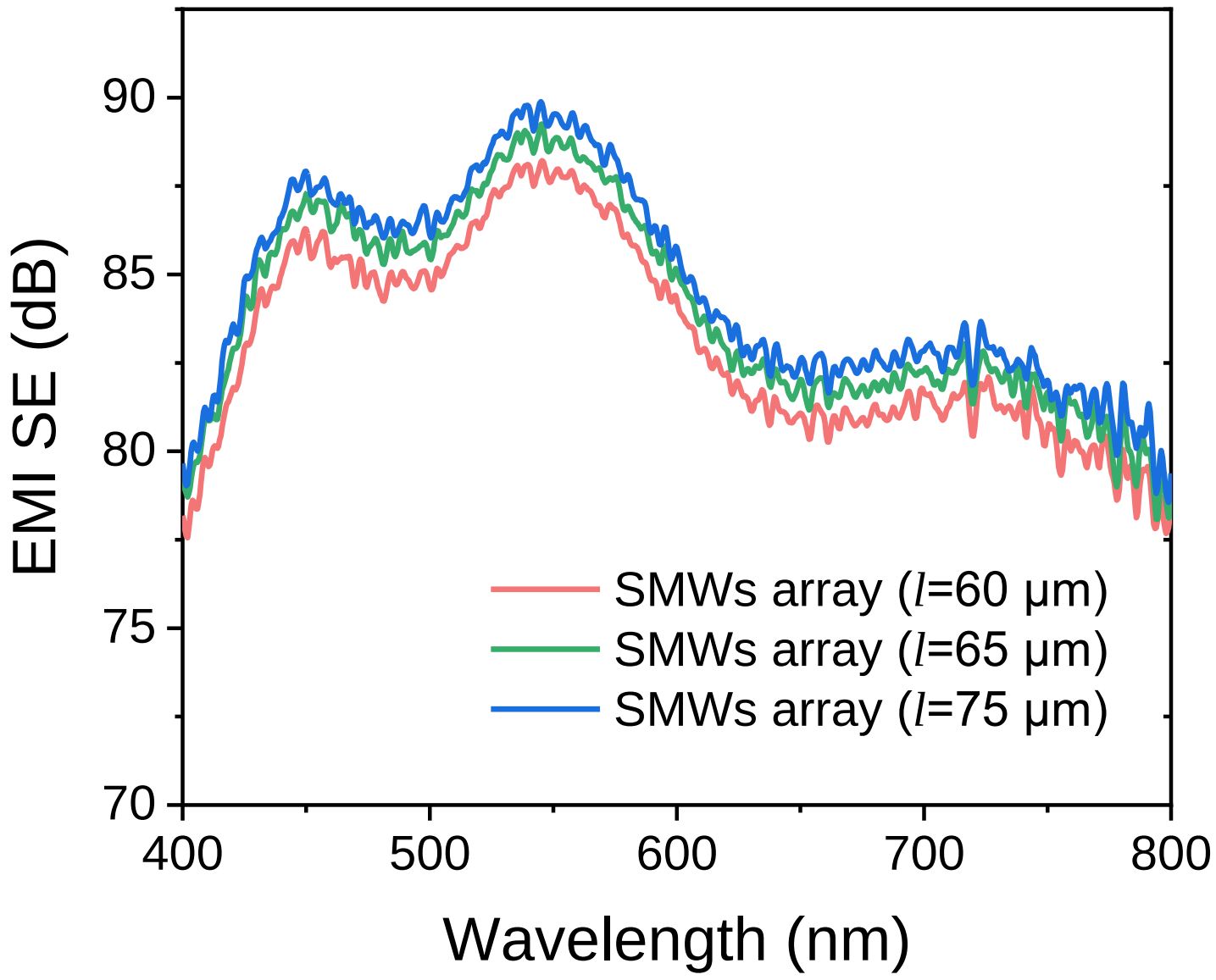

**Fig. S2 Optical transmittances of SMWs array with different structural parameters**.

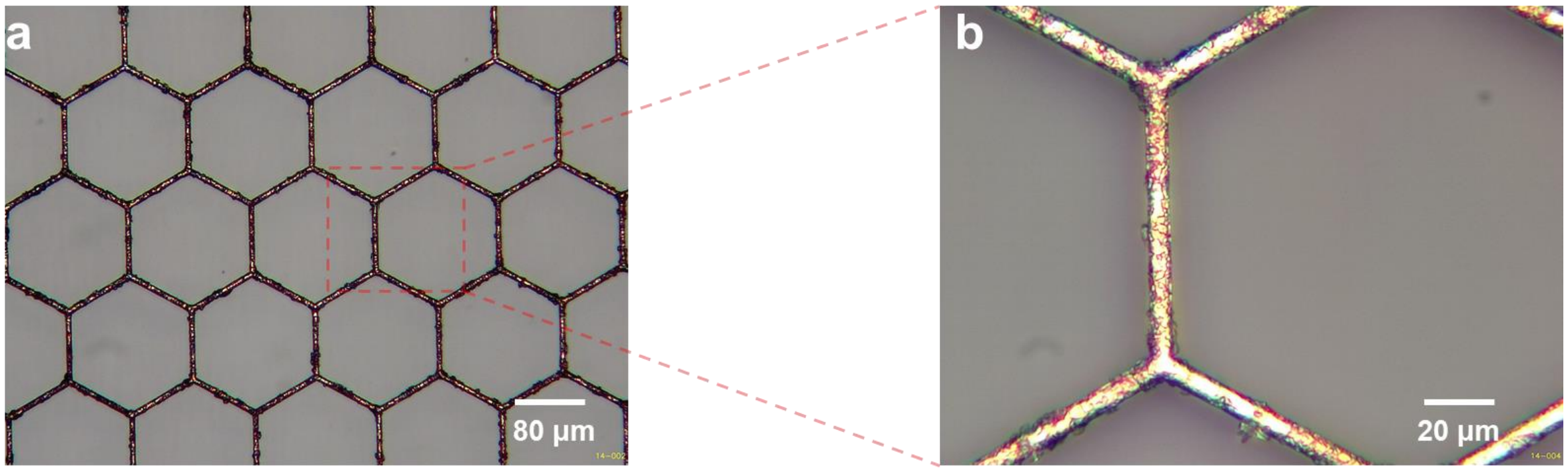

**Fig. S3 Micrograph of apertures under different magnification**. **a** 100X and **b** 400X.

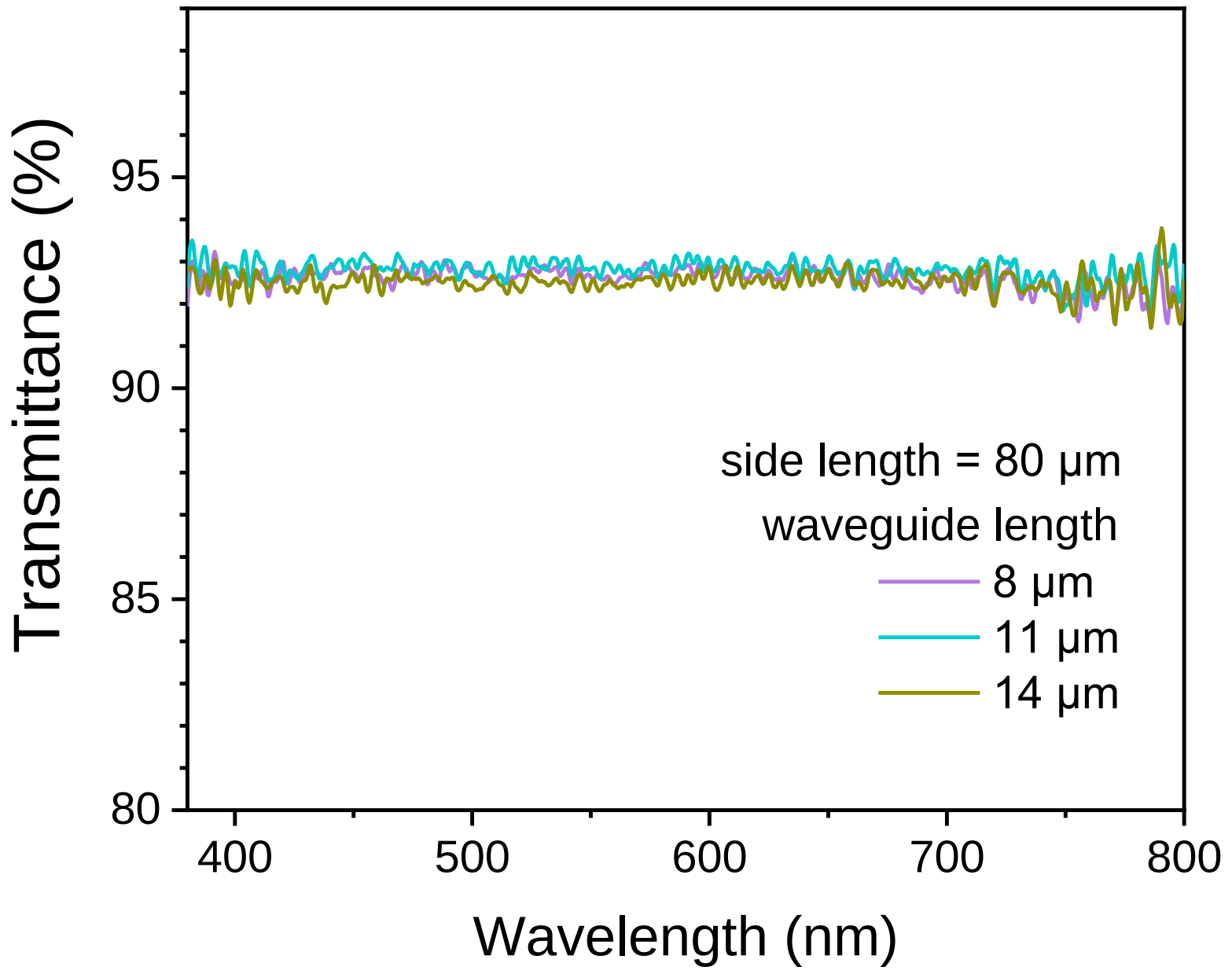

**Fig. S4 Optical transmittance of freestanding micro-waveguides (side length = 80 μm) with different waveguide lengths.** Since there is no change in light transmission aperture, they exhibit almost identical light transmission.

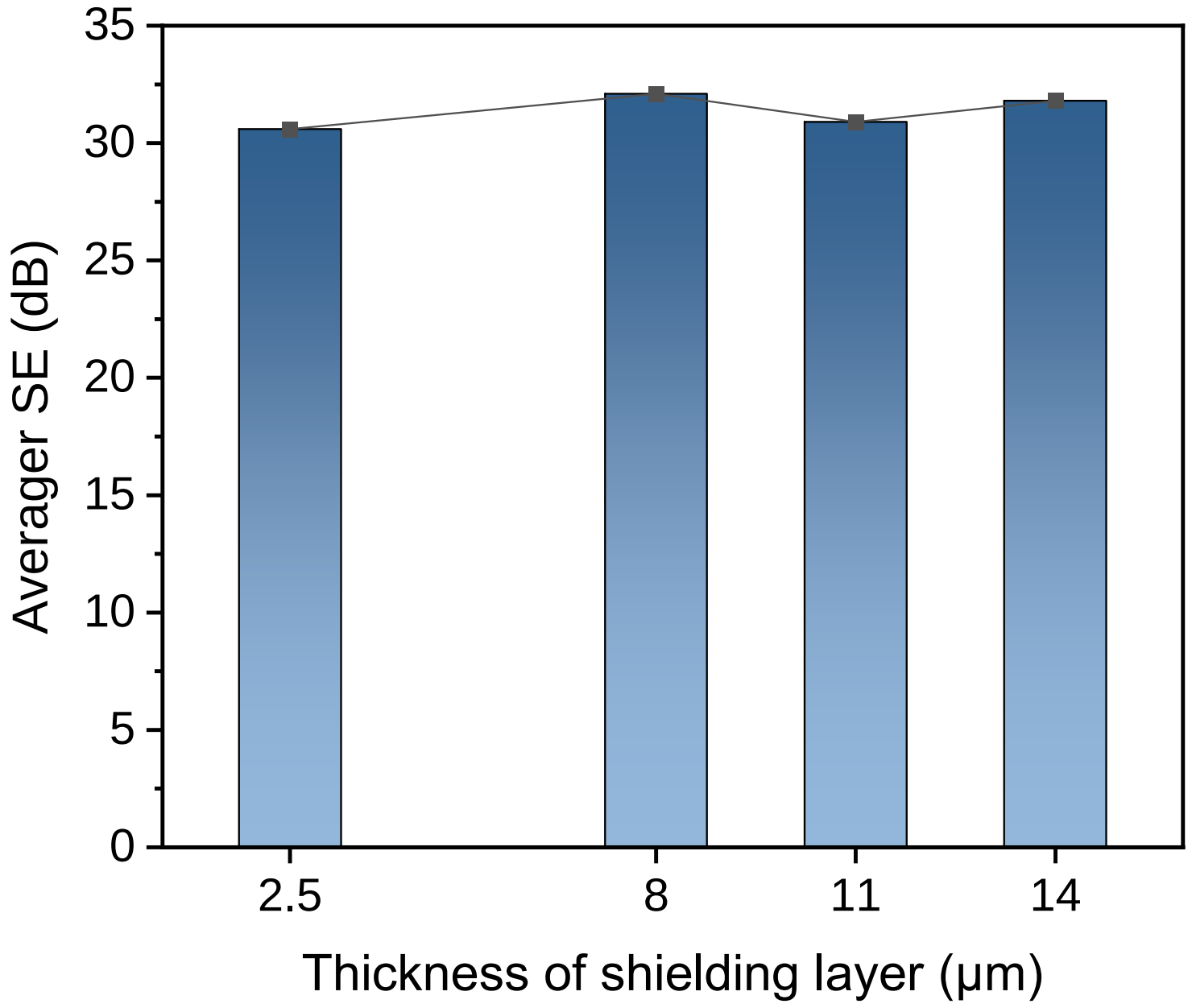

**Fig. S5** Average shielding performance of thin metal mesh (2.5 μm in thickness) and freestanding micro-waveguide layers with varying waveguide lengths (8 μm, 11 μm, 14 μm) in X-band. The side lengths of apertures are 80 μm.

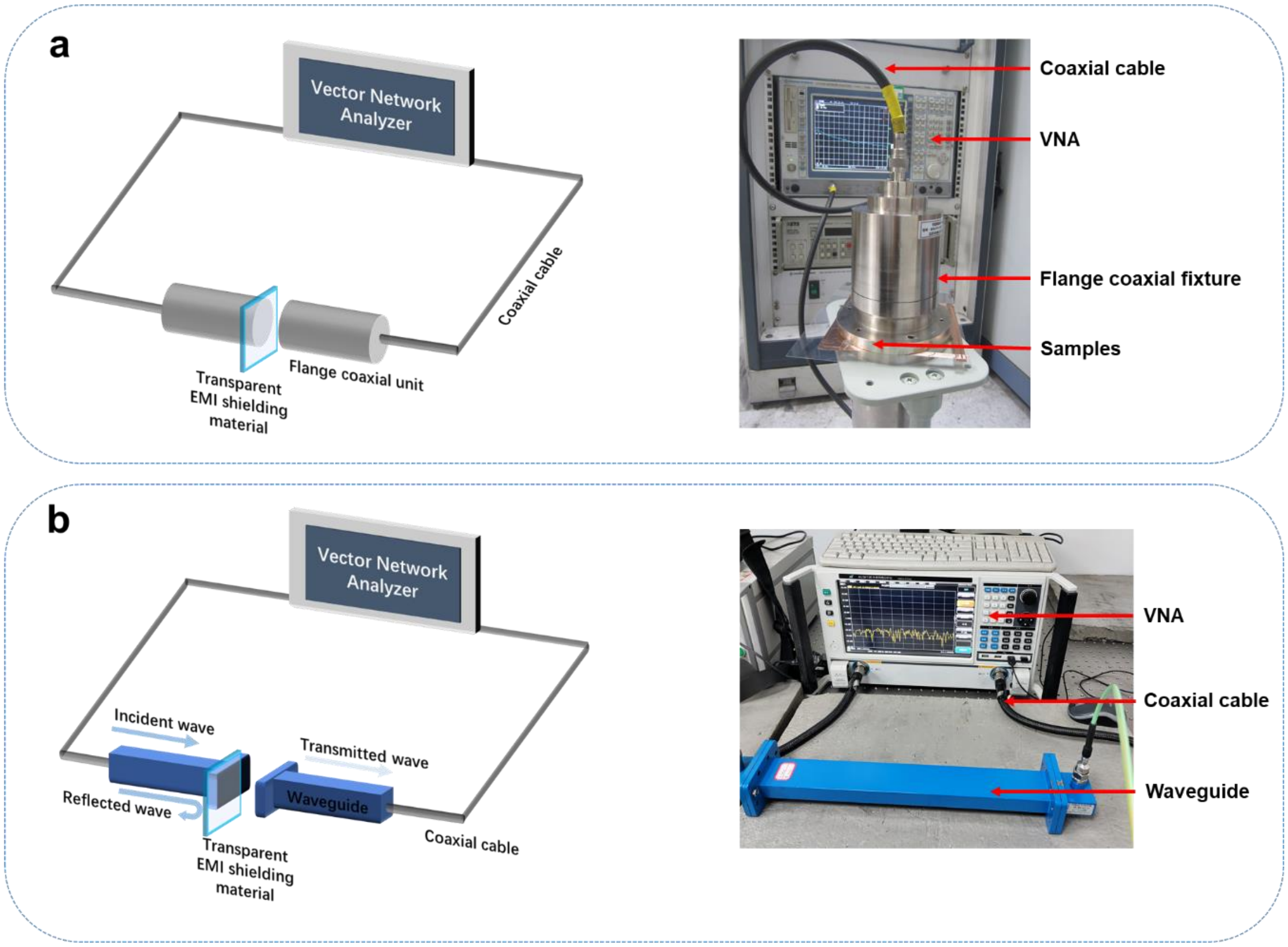

**Fig. S6 EMI SE measurement set-up. a** Schematic diagram and the experimental setup of the Flanged coaxial EMI SE measurement fixture used in the frequency range from 0.03 GHz to 1.5 GHz. **b** Waveguide measurement system employed in the frequency range from 1.7 GHz to 18 GHz. We used six sets of waveguides operated in various frequency bands (1.7–2.6 GHz, 2.6–4 GHz, 4–6 GHz, 6–8 GHz, 8–12 GHz, and 12–18 GHz) to obtain data over broadband frequencies.

**Supplementary Table S1**. **Comparison of figures-of-merits for different transparent EMI shielding materials**

| Materials and Structures | Transmittance (%) | SE (dB) | Thickness (μm) | Frequency range (GHz) | Ref. |
|---|---|---|---|---|---|
| PEDOT:PSS | 97 | 15 | 0.05 | 8–12 | *ACS Appl. Mater. Interfaces* **2020**, 12, 28596 |
| CA/AgNW/PU | 92,<br>81 | 20.7,<br>30.3 | —<br>— | 8–12 | *ACS Appl. Mater. Interfaces* **2018**, 10, 11941 |
| A/RGO/SANW<br>A/GO/SANW | 85<br>85 | 24<br>28 | —<br>— | 0.03–3 | *ACS Appl. Mater. Interfaces* **2018**, 10, 29730 |
| AgNW/PEDOT:PSS composite film | 91<br>81.1 | 30.5<br>41.4 | 0.1 | 1–12 | Macromol. Mater. Eng. **2020**, 2000607 |
| ITO/Cu-doped Ag/ITO | 96.5 | 26 | 0.088 | 8–40 | *ACS Appl. Mater. Interfaces* **2019**, 11, 11782 |
| $Fe_3O_4$/PVP modified Ag NWs | 90 | 24.9 | — | 8–12 | *ACS Appl. Mater. Interfaces* **2020**, 12, 2826 |
| Ag NW/PDDA | 91.3<br>86.8 | 28<br>31.3 | —<br>— | 8–12 | *Nanoscale* **2020**, 12, 14589 |
| MXene/AgNW-PVA | 52.3 | 32 | <10 | 8–12 | *ACS Appl. Mater. Interfaces* **2020**, 12, 40859 |
| AgNW@rGO | 91.1<br>61.2 | 35.5<br>57.6 | —<br>— | 8–12 | *ACS Nano* **2020**, 14, 8754 |
| Steel Fiber/Silicone Resin | 70<br>67 | 15-26<br>27-36 | —<br>— | 0.3–1.5 | *IEEE Trans. Electromagn. Compat.* **2014**, 56, 328 |
| UVA-illuminated AgNW films | 90 | 25dB | — | 8–12 | *Adv. Mater. Interfaces* **2018**, 6, 1801635 |
| AZO (45 nm)/Ag (9 nm)/AZO (45 nm) | ~90 | ~39 | 0.099 | 1.5–3 | *Current Applied Physics* **2016**, 16, 1642 |
| graphene/50 nm Ag-mesh film | 90 | 14.1 | — | 12–18 | *Carbon* **2017**, 115, 34 |
| Ag mesh | 88 | 23 | — | 1.5–10 | *Curr. Appl. Phys.* **2019**, 19, 8 |
| Ni mesh | 92 | 40 | 2.5-6.0 | 8–12 | *Opt. Express* **2019**, 27, 24194 |
| Crackle metallic mesh | 91 | 26 | 3 | 12–18 | *Sci. Rep.* **2016**, 6, 25601 |
| Cu–Ag mesh | 85.4<br>82.2 | 42.4<br>43.7 | 0.289 | @8 | *J. Mater. Sci.* **2021**, 56, 14741 |
| Graphene/metallic mesh | 91<br>91 | 23.60-28.91<br>13.48-14.52 | —<br>— | 12–18<br>26.5–40 | *ACS Appl. Mater. Interfaces* **2017**, 9, 34221 |
| graphene/Ni mesh | 83 | 12.1 | 0.020 | 0.75–3 | *Acs Applied Nano Materials* **2020**, 3, 7474 |
| Cu mesh | 82 | 24 | 2 | 12–18 | *Adv. Electron. Mater. **2018**, 1800156* |
| Metallic mesh | 68.4 | 31.4 | 2 | @2.45 | *IEEE Trans. Electromagn. Compat.* **2020**, 62, 1076 |
| Layered graphene/PET | 80.5 | 19.14 | 8000 | 18–26.5 | *Nanoscale* **2016**, 8, 16684 |
| PEI/RGO<br>Double- PEI/RGO | 73<br>62 | 3.09<br>6.37 | 2.02<br>40.04 | 0.5–8.5 | *ACS Appl. Mater. Interfaces* **2014**, 6, 17647 |
| Double-layer metal mesh separated by glass | 85.73<br>76.35 | 19.74–48.34<br>37.61-75.84 | 6001.2 | 0.15–5 | *RSC Adv.* **2019**, 9, 22282 |
| Double-layer multi-ring metallic mesh | ~90 | 27-37 | 800.4 | — | *Opt. Lett.* **2017**, 42, 1620 |
| 2-graphene-2-mesh | 85 | 47.79–67.92 | 4800 | 12–18 | *2D Mater.* **2017**, 4, 025021 |
| Layered Ti3C2Tx MXene /AgNW | 83 | 49.2 | ~18 | 8–12 | *ACS Nano* **2020**, 14, 16643 |
| ZnO/Ag/ZnO<br>Double-ZnO/Ag/ZnO<br>Double-ZnO/Ag/ZnO with<br>0.2mm glass<br>-1.1mm glass<br>-1.5mm glass | 91.89<br>80.36<br>88.91<br>88.26<br>89.03<br>86.69 | 34.73<br>39.81<br>40.16−56.34<br>55.2-70.76<br>48.94-68.78<br>53.31-67.70 | 0.091<br>0.182<br>~200<br>~1100<br>~1500<br>~2000 | 4–40 | *ACS Appl. Mater. Interfaces* **2020**, 12, 26659 |
| -2mm glass | | | | | |
| Metal mesh film (MMF) | 70 | 19 | 5 | 8–12 | *Sci. Rep.* **2021**, 11, 16161 |
| MMF/glass/saltwater/glass | 64 | 61 | 14005 | | |
| MMF/glass/saltwater/glass/MMF | 45 | 82 | 14010 | | |
| **Shorted micro-waveguides array** | 90.4 | 62.2 | ~1014 | 1.7–18 | This work |

**References**

1 Peng, H., Zhiguo, L. & Jiancheng, Q. Simulation Analysis of Electromagnetic Shield-ing Effectiveness in Ventilation Window of Waveguide. *J. Simul.* **5**, 93 (2017).

2 Wang, H. *et al.* Highly Transparent and Broadband Electromagnetic Interference Shielding Based on Ultrathin Doped Ag and Conducting Oxides Hybrid Film Structures. *ACS Appl. Mater. Interfaces* **11**, 11782-11791, (2019).

3 Macleod, H. A. & Macleod, H. A. *Thin-film optical filters*. CRC Press: Tucson, Arizona, 2010.